\documentclass[12pt]{article}
\usepackage{graphicx}
\usepackage{amssymb}

\usepackage{scicite}

\usepackage{times}

\newenvironment{sciabstract}{%
\begin{quote} \bf}
{\end{quote}}

\newcounter{lastnote}

\title{A circumbinary debris disk\\ in a polluted white dwarf system}

\author
{J. Farihi$^{1,5\ast}$, S. G. Parsons$^{2,3}$, B. T. G\"ansicke$^{4}$\\
\\
\normalsize{$^{1}$Department of Physics and Astronomy, University College London, London WC1E 6BT, UK}\\
\normalsize{$^{2}$Departamento de F\'isica y Astronom\'ia, Universidad de Valpara\'iso, Valpara\'iso 2360102, Chile}\\
\normalsize{$^{3}$Department of Physics and Astronomy, University of Sheffield, Sheffield S3 7RH, UK}\\
\normalsize{$^{4}$Department of Physics, University of Warwick, Coventry CV5 7AL, UK}\\
\normalsize{$^{5}$STFC Ernest Rutherford Fellow}\\
\\
\normalsize{$^\ast$To whom correspondence should be addressed; E-mail:  j.farihi@ucl.ac.uk}
}

\date{}

\begin{document} 


\baselineskip24pt


\maketitle 


\begin{sciabstract}

Planetary systems commonly survive the evolution of single stars, as evidenced by terrestrial-like planetesimal debris observed 
orbiting and polluting the surfaces of white dwarfs \cite{jur14,far16}.  This letter reports the identification of a circumbinary dust disk 
surrounding a white dwarf with a substellar companion in a 2.27\,hr orbit.  The system bears the dual hallmarks of atmospheric metal 
pollution and infrared excess \cite{gir11,far12}, however the standard (flat and opaque) disk configuration is dynamically precluded 
by the binary.  Instead, the detected reservoir of debris must lie well beyond the Roche limit in an optically thin configuration, where 
erosion by stellar irradiation is relatively rapid.  This finding demonstrates that rocky planetesimal formation is robust around close 
binaries, even those with low mass ratios.

\end{sciabstract}

\clearpage 

The formation and evolution of planetary systems around close binary stars is a challenging problem, yet provides insight into 
the growth of planetesimals within evolving protoplanetary disks and planet formation in general.  The small but increasing 
number of transiting circumbinary planets detected with {\em Kepler} \cite{doy11,wel12} are providing the first tests of theoretical 
formation models.  To date, it has been shown that in situ formation is unfavorable for most of these Neptune- to Jupiter-sized 
bodies due to the destructive, dynamical effects of the central binary on planetesimal agglomeration in the regions where the 
planets currently orbit \cite{paa12,pel13,lin14}.  However, for a range of masses including small planets, recent models predict 
favorable conditions for planetesimal growth within the snow line, and thus promoting efficient terrestrial planet formation around 
binaries \cite{raf13,mar13,bro15}.

Atmospheric pollution in white dwarf stars offers a unique and powerful window into the assembly and chemistry of terrestrial 
exoplanets.  There are now more than three dozen planetary system remnants made evident based on thermal and line emission 
from circumstellar disks \cite{far16}, and several hundred where photospheric metals indicate ongoing or recent accretion of 
planetary debris \cite{koe14}.  The current paradigm of disrupted and accreted asteroids has been unequivocally confirmed 
by numerous studies, including the recent detection of complex and rapidly evolving photometric transits from debris fragments
orbiting near the Roche limit of one star \cite{van15,xu16,gan16}.  To date, all polluted white dwarfs with detailed analyses 
indicate the sources are rocky planetesimals comparable in both mass and composition to large Solar System asteroids 
\cite{gan12,far13,jur14}, and thus objects that formed within a snow line.  These findings unambiguously demonstrate that 
large planetesimal formation in the terrestrial zone of stars is robust and common.

\begin{figure}[ht!]
\begin{center}
\begin{tabular}{c c}
\hskip -5pt
\includegraphics[height=2.5in]{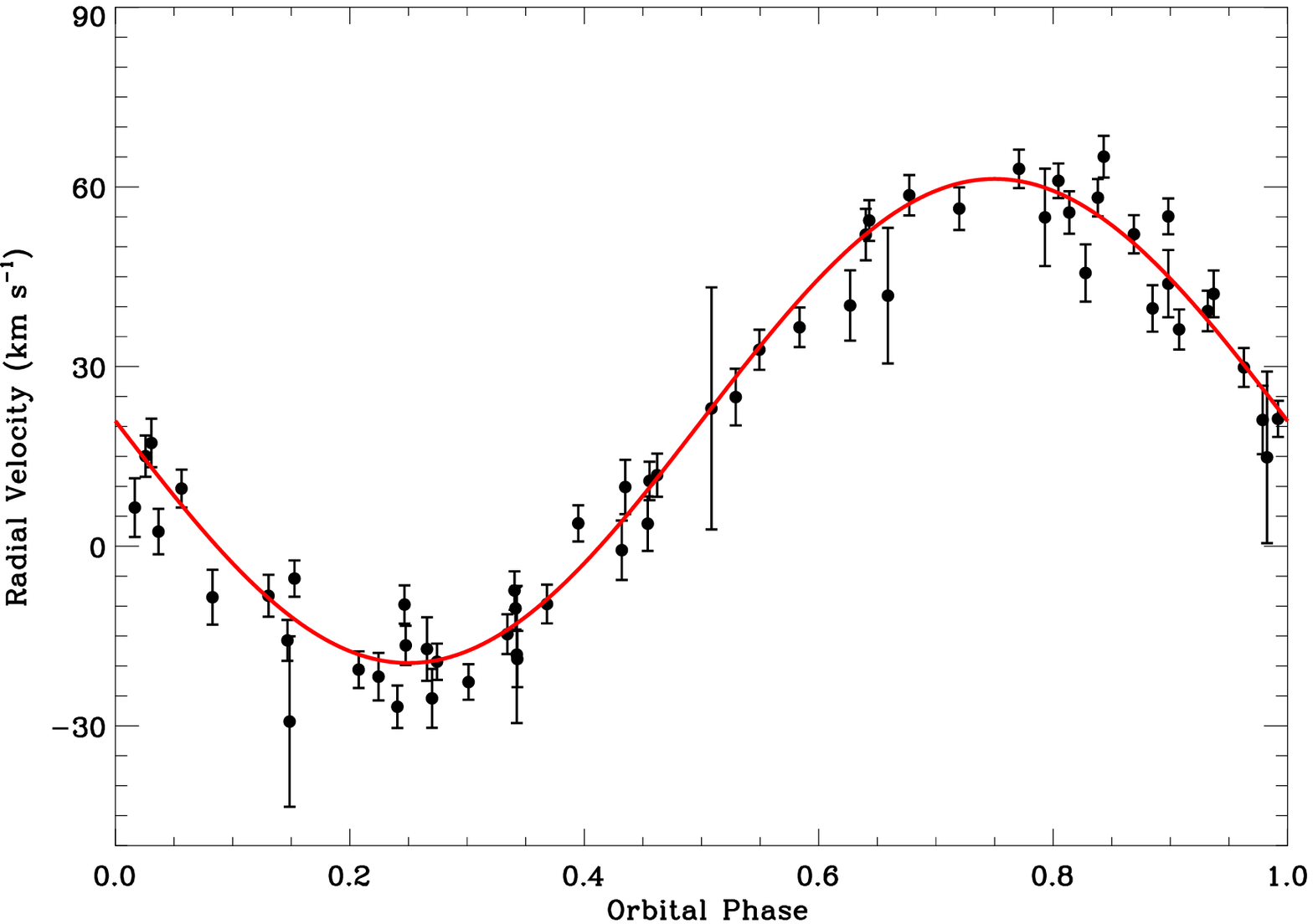} & \hskip 0pt
\includegraphics[height=2.5in]{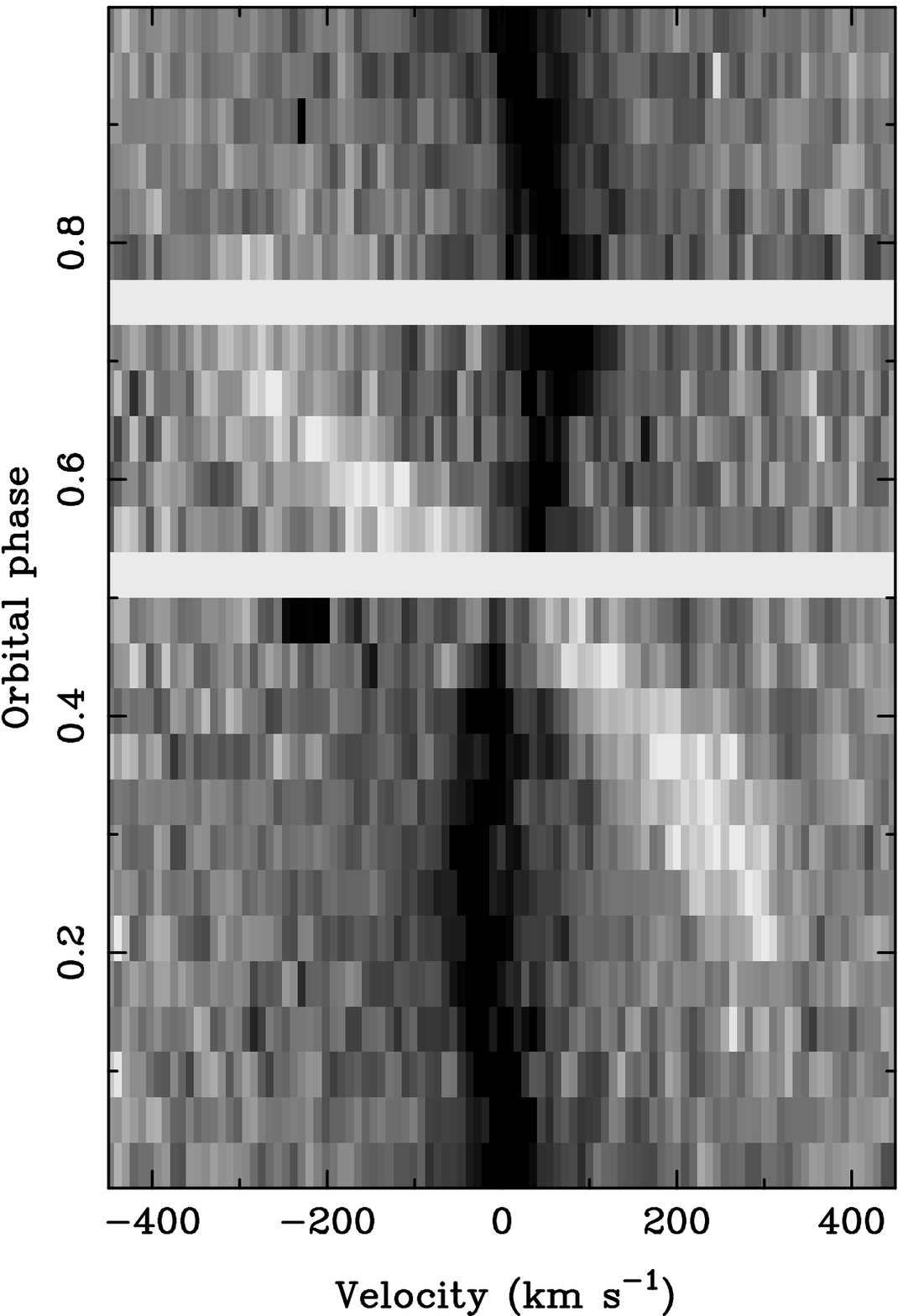}\\
\end{tabular}
\end{center}
\vskip -5 pt
{\footnotesize {\bf Figure 1: Phased radial velocity diagrams.}  The left panel plots the phased radial velocities for SDSS\,1557 based on 
measurements of the Mg\,{\sc ii} absorption line and their associated errors (see Methods), with the resulting parameters 
given in Table 1.  The right panel shows a trailed spectrum for the H$\alpha$ component of SDSS\,1557, where darker 
features represent lower fluxes and lighter features indicate higher fluxes.  The absorption component follows identically 
that of Mg\,{\sc ii}, and thus both features are intrinsic to the white dwarf photosphere.  The emission component is observed 
in anti-phase with the primary, and therefore originates from the irradiatively heated companion.  The two horizontal blank 
lines represent phase bins without coverage.}  
\end{figure}

Until now, over 90\% of such dusty and polluted white dwarf systems have been discovered among single stars, with a small 
fraction belonging to sufficiently wide binaries ($a\gg100$\,AU) where the evolution of each star -- and any associated planetary
system -- proceeds as a singleton.  The white dwarf SDSS\,J155720.77+091624.6 (hereafter SDSS\,1557) exhibits strong infrared 
excess from $T\approx1100$\,K dust, and possesses one of the highest known metal abundances \cite{gir11,far12}, but has a 
distinctive low mass ($M\lesssim0.45\,M_{\odot}$) that indicates a helium core.  Such white dwarf masses are too low for single 
stars to attain within the age of the Galaxy, implying SDSS\,1557 belongs to a class of remnants whose evolution was truncated 
prior to helium ignition \cite{swe94,fon01}.  These low-mass, helium core white dwarfs are often found to be short-period, 
spectroscopic binaries \cite{mar95}, consistent with efficient (common) envelope ejection during the first ascent giant branch.

\begin{table}
\begin{center}
\small	
\begin{tabular}{@{}lrlr@{}}

\hline

\multicolumn{2}{c}{White Dwarf Primary}	&\multicolumn{2}{c}{Binary}\\

\hline

								&					&$p$ (h)							&$2.273153\pm0.000002$\\
SpT								&DAZ				&$t_0$ (HJD)						&$2457201.0551\pm0.0004$\\
$g$ (AB\,mag)						&18.43				&$K_1$ (km\,s$^{-1}$)				&$40.42\pm0.69$\\
$T_{\rm eff}$ (K)					&$21\,800\pm800$		&$K_2$ (km\,s$^{-1}$)				&$288.3\pm3.0$\\
$\log\,g$ (cm\,s$^{-2}$)				&$7.63\pm0.11$		&$\gamma_1$ (km\,s$^{-1}$)			&$20.91\pm0.51$\\
$M_1$ ($M_{\odot}$)				&$0.447\pm0.043$		&$\gamma_2$ (km\,s$^{-1}$)			&$6.2\pm1.4$\\
$\log(L/L_{\odot})$					&$-1.25\pm0.08$		&$z$ (km\,s$^{-1}$)					&$14.7\pm1.5$\\
Cooling Age (Myr)					&$33\pm5$			&$M_2$ ($M_{\odot}$)				&$0.063\pm0.002$\\
$d$ (pc)							&$520\pm35$			&$i$ ($^{\circ}$)					&$62\pm3$\\
								&					&$a$ ($R_{\odot}$)					&$0.70\pm0.02$\\

\hline

\end{tabular}
\end{center}
\normalsize

{\footnotesize {\bf Table 1: Stellar and binary parameters.}  White dwarf parameters $T_{\rm eff}$ and $\log\,g$ were derived from 
fitting atmospheric models to the higher Balmer lines for individual X-shooter spectra, and the adopted errors are the standard 
deviation from all 60 measurements.  Errors in derived stellar parameters were calculated by propagating the uncertainties 
in the adopted $T_{\rm eff}$, $\log\,g$, and published photometry through white dwarf evolutionary models \cite{fon01}.  The 
binary and companion parameters, including errors, were calculated from analysis of the radial velocity measurements of the 
Mg\,{\sc ii} line in all X-shooter spectra (see Methods), and Kepler's laws.  Note $t_0$ is defined as the inferior conjunction of 
the white dwarf, and $K_2$ is calculated assuming the observed emission is uniformly spread across the inner hemisphere 
of the companion.}

\end{table}

SDSS\,1557 was observed on multiple occasions with the GMOS and X-Shooter spectrographs at Gemini Observatory South 
and the Very Large Telescope respectively.  The lower resolution GMOS data hinted at radial velocity changes in the Mg\,{\sc ii} 
4482\,\AA \ absorption feature, and the higher resolution X-shooter data set reveal a robust period of 2.273\,hr with semi-amplitude 
$K_1=40.4$\,km\,s$^{-1}$ (Figure 1).  A close examination of the H$\alpha$ region reveals an emission feature arising from the 
irradiated companion, seen in anti-phase and with semi-amplitude of 267.6\,km\,s$^{-1}$.  Assuming the emission is uniformly 
spread across the inner hemisphere of the companion, this implies a binary mass ratio $M_2/M_1=0.14$.  Both the radial 
velocity data and the spectral energy distribution are consistent with a substellar companion of $0.063\,M_{\odot}$ ($66\,M
_{\rm Jup}$) seen at an orbital inclination near 63$^{\circ}$. Table 1 lists all system parameters with uncertainties.

\begin{figure}[ht!]
\begin{center}
\includegraphics[height=3.0in]{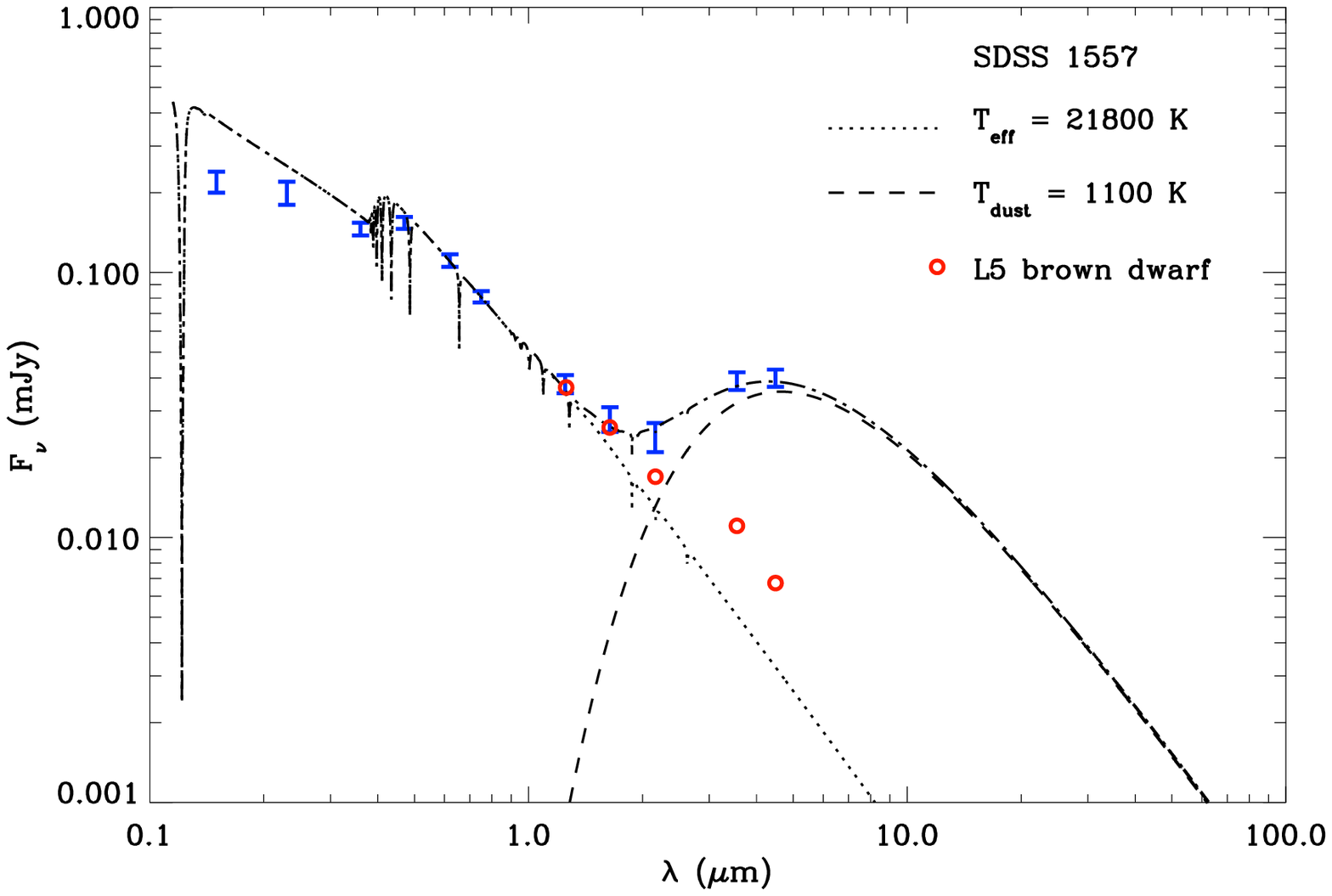}
\end{center}
\vskip -15 pt
{\footnotesize {\bf Figure 2: Spectral energy distribution.}  Previously reported multi-wavelength photometric data for SDSS\,1557 are plotted 
as blue error bars \cite{far12}, with the stellar flux from a pure hydrogen atmospheric model is shown as a dotted line.  Overplotted 
as a dashed line is an 1100\,K blackbody fitted to the strong, infrared dust emission, and an L5-type brown dwarf is shown as red 
circles; any companion earlier than L3 is ruled out by the photometry below 2\,$\mu$m.  The measured 4.5\,$\mu$m flux from the 
system is between 5 and 6 times brighter than any allowed companion, and hence must be due to circumbinary dust.  There is a 
notable deficit in ultraviolet flux relative to the stellar model, but a metal-rich atmosphere cannot account for the mismatch; neither 
the photospheric opacity nor the emergent flux are strongly affected by the presence of trace metals in a hydrogen-rich star of this 
temperature.  The presence of a vertically extended dust shell may account for additional extinction at these shortest wavelengths.}  
\end{figure}

Figure 2 shows the expected infrared emission from a mid-L type substellar companion at the white dwarf distance based on 
the derived stellar parameters.  Ultracool dwarfs earlier than L3 can be firmly ruled out, as the expected fluxes blueward of 
2\,$\mu$m would be significantly brighter than those observed.  At the same time, the substellar companion cannot account 
for the total excess emission in the system over the $2-5\,\mu$m region, as the 4.5\,$\mu$m flux is 5 to 6 times larger than 
expected for an L3 or L5 type dwarf \cite{pat06}.  Even in the extreme case where the predicted 4.5\,$\mu$m flux from a mid-L 
type companion is nearly doubled -- mimicking the peak-to-peak changes seen in related systems due to irradiation from the 
primary \cite{cas15} -- the observed flux remains significantly higher.  SDSS\,1557 thus has both a strong infrared excess 
and atmospheric metal pollution at high abundance.  It is therefore similar to more than three dozen known white dwarfs 
that are accreting planetary debris from circumstellar reservoirs that are consistent with tidally disrupted minor planets.

The identification of orbiting dust and atmospheric pollution within a close binary presents a fundamental challenge to the disk 
modeling, as the canonical flat and opaque ring geometry is dynamically prohibited.  This standard disk would nominally be 
completely contained between 0.4 and 0.9\,$R_{\odot}$ from SDSS\,1557 \cite{far12}.  However, for a non-eccentric orbit, a 
companion of mass $M_2=0.063\,M_{\odot}$, and a semimajor axis $a=0.70\,R_{\odot}$, stable {\em circumstellar} orbits are 
allowed within $0.4a\approx0.3\,R_{\odot}$ of the primary, and stable {\em circumbinary} orbits are allowed beyond $2.0a\approx
1.4\,R_{\odot}$ \cite{hol99}.  The emitting dust cannot lie interior to $0.3\,R_{\odot}$, where even shielded grains would attain 
$T>1700$\,K and thus be inconsistent with the observed thermal emission.  Therefore the solid material must lie exterior to 
$1.4\,R_{\odot}$, but at this distance a flat, opaque disk would be no warmer than 500\,K, and by itself inconsistent with the 
infrared data.  Blackbody grains with $T=1100$\,K will have an orbital radius of $3.3\,R_{\odot}$ in this system; this is the only 
configuration consistent with both the observed emission and the dynamical effects of the binary.  Figure 3 provides an illustration 
of the system.

This $T_{\rm eff}=21\,800$\,K hydrogen-rich white dwarf has a heavy element sinking timescale of only a few weeks \cite{koe09}, 
and observations of the Mg\,{\sc ii} feature include detections in 2010 April, 2012 May, 2014 May, and 2016 July, with no observed 
change in line strength.  Thus SDSS\,1557 is in a steady state of accretion from the circumbinary disk; the inferred accretion 
rate is $\dot M \simeq 6 \times 10^8$\,g\,s$^{-1}$ assuming a bulk Earth abundance for Mg.  Captured material from the 
companion -- with solar abundance Mg -- can be dismissed as the source of atmospheric enrichment, since this would require 
the brown dwarf to either 1) have a wind rate higher than the solar value \cite{deb06}, or 2) be twice its normal size to overfill 
its Roche lobe.  It is noteworthy that the similar and even closer binary WD\,0137--349 neither exhibits atmospheric metals nor 
an infrared excess above that expected for its components \cite{max06}.

\begin{figure}
\begin{center}
\begin{tabular}{l l}
\hskip -25pt
\includegraphics[height=3.0in]{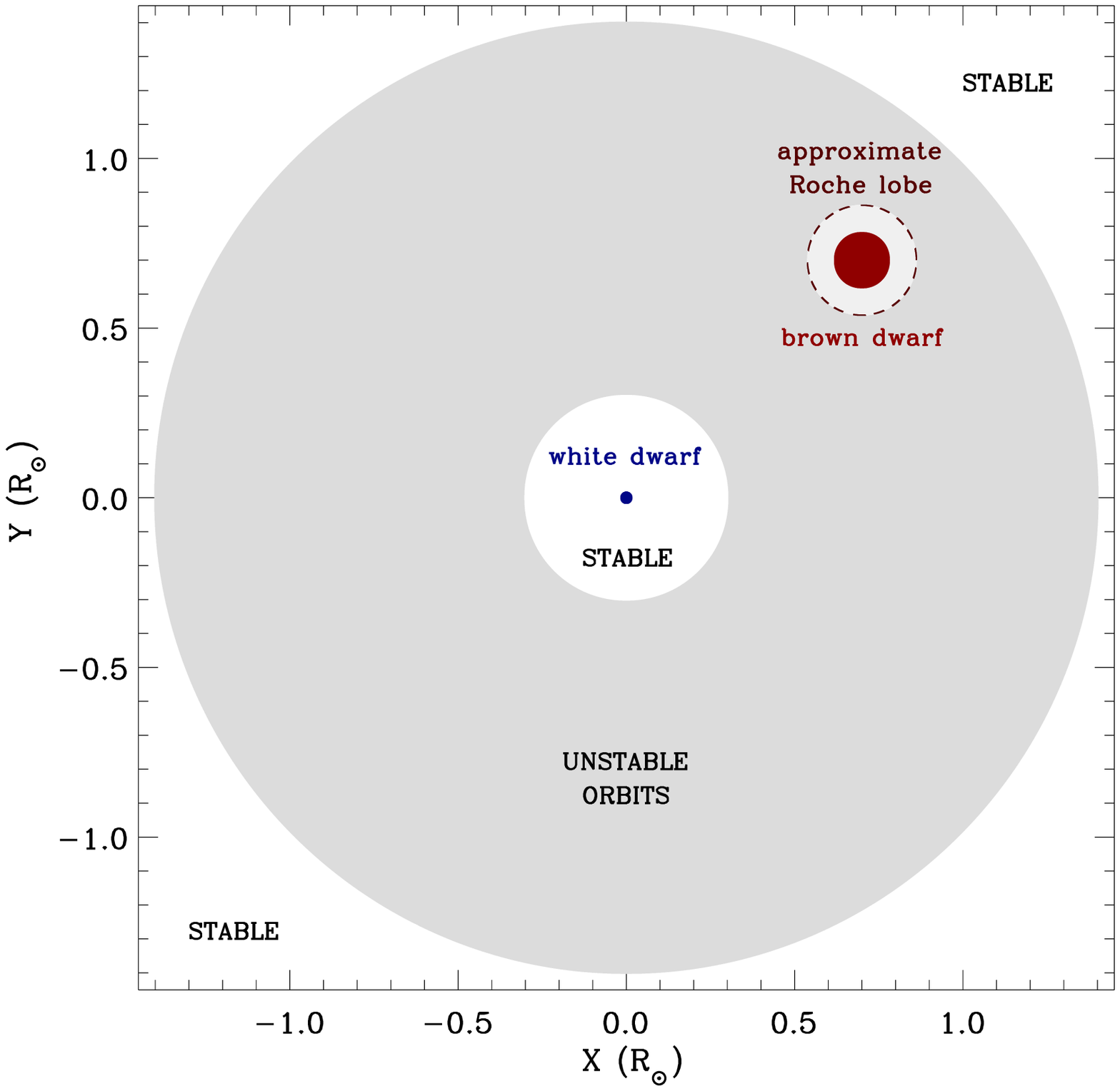} & \hskip -5pt
\includegraphics[height=3.0in]{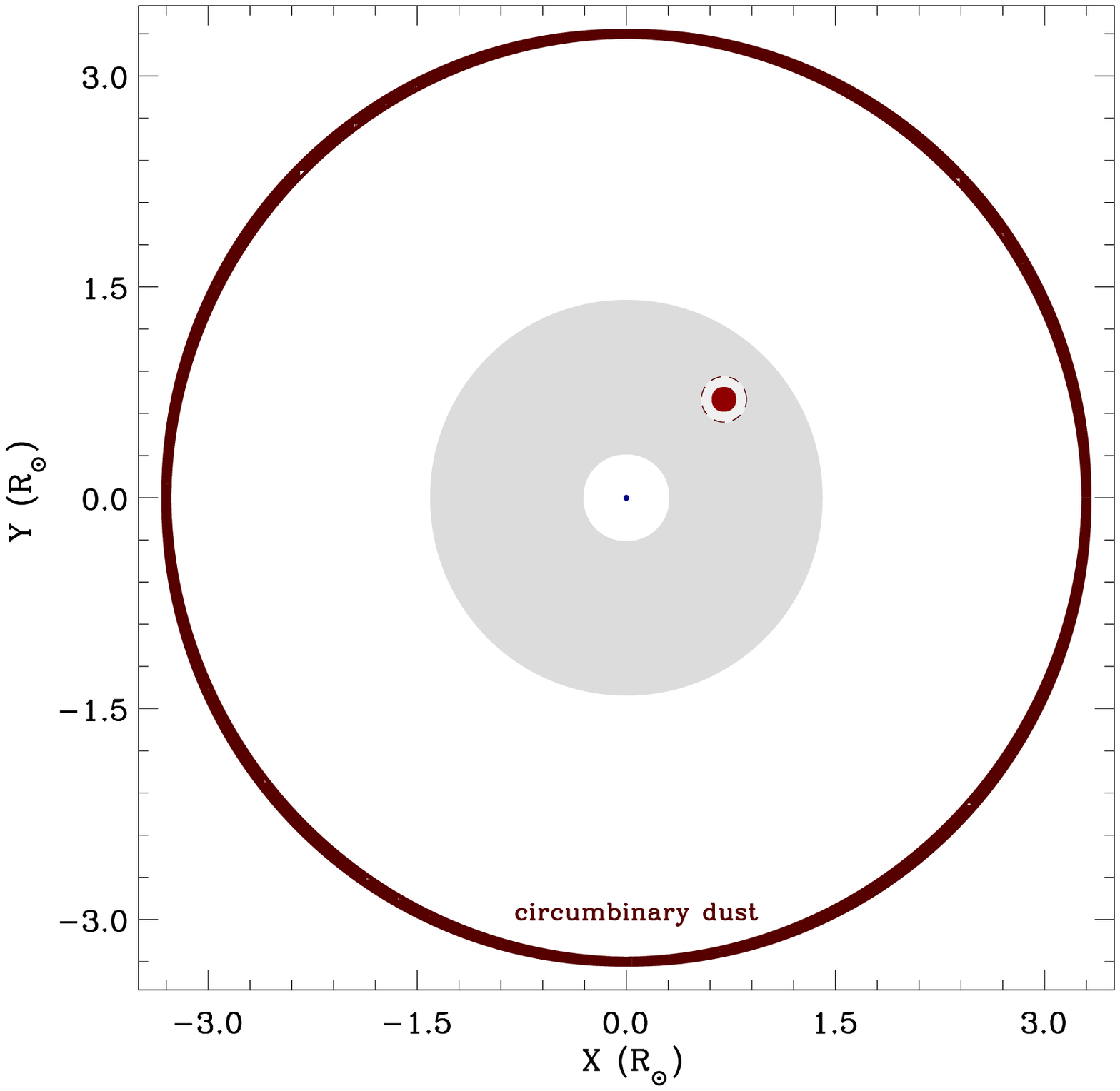}\\
\end{tabular}
\end{center}
\vskip -10 pt
{\footnotesize {\bf Figure 3: Circumbinary system geometry.}  Diagrams illustrating SDSS\,1557 in the primary reference frame, where the entire 
system is drawn within a single plane as viewed from above, with the stable and unstable orbital regions \cite{hol99} labelled on the 
left.  The sizes of the binary components are accurately portrayed in both diagrams, and the companion is surrounded by a dashed 
boundary delineating a circular approximation of its Roche lobe.}
\end{figure}

This is the first case of dusty debris inferred to be exterior to the Roche limit of a metal-accreting white dwarf \cite{jur07,far09}.  
The surrounding reservoir was likely created via binary driving of debris generated during a catastrophic fragmentation 
interior to the Roche limit.  Material must flow onto the stellar surface via stable streaming from the inner edge of the circumbinary 
disk, where solids will gradually sublimate as they spiral inward from $3.3\,R_{\odot}$.  This process is well understood and likely 
detected in young binaries \cite{art96,ter15}.

The lifetime of an optically thin dust reservoir is determined by Poynting-Robertson (PR) drag, scaling linearly with particle 
size and yielding 9\,yr for micron-sized grains orbiting SDSS\,1557 at $3.3\,R_{\odot}$.  Therefore, the circumbinary dust disk 
surrounding -- and polluting -- the white dwarf is either 1) in a transient phase of rapid erosion, 2) being replenished by additional
dust sources, or 3) contains a significant mass of particles larger than 10-30\,$\mu$m.  In the second case, the observed 1100\,K
dust shell could be fed by an exterior, flat and cold disk that is beyond the gravitational influence of the binary.

Models of optically thin disk accretion onto single white dwarfs predict rates two orders of magnitude lower than that inferred to 
be ongoing in this system \cite{boc11}, where those calculations were made for vertically narrow (i.e.\ flat) disks, as expected from 
rapid dynamical relaxation around single stars.  The accretion rate in this system therefore suggests the disk must have a significant 
vertical extent.  As PR drag drives particles inward towards SDSS\,1557, they will eventually experience gravitational encounters 
with the binary, and this could produce and maintain a vertically thick yet optically thin, disk configuration.  Comparing the current,
estimated rate of accretion with the theoretical maximum for PR drag on optically thin dust, implies the disk intercepts a fraction 
$\tau=0.0025$ of the total starlight.  However, this is only about half of $L_{\rm IR}/L_*=0.0055$ obtained from fitting the spectral 
energy distribution, and may imply the accretion rate is actually more than twice the estimate based on models.  

A cylindrical dust shell with radius $3.3\,R_{\odot}$ and height $1.1\times10^9$\,cm (1.0 white dwarf radius) would cover a fractional 
area of 0.0025.  Such a shell might account for the observed ultraviolet fluxes shown in Figure 2, which sit below the atmospheric 
model and cannot be adequately explained by interstellar reddening, as this would result in a significantly lower effective temperature 
than derived here and by other authors \cite{gir11}.  If correct, ultraviolet spectroscopy has the potential to detect gas absorption 
along the line of sight through the shell.

The progenitors of extant white dwarfs are typically A- and F-type stars with masses in the range $1.2-3.0\,M_{\odot}$, a stellar
population that is not readily probed for inner planetary systems using conventional techniques.  The SDSS\,1557 binary was thus 
born with a low mass ratio in the range $M_2/M_1=0.02-0.05$, and likely with a semimajor axis less than an AU or so, in order to 
form a common envelope on the first ascent giant branch.  Excepting the low mass ratio, the progenitor binary would have been 
similar to the eclipsing binaries with planets detected by {\em Kepler} \cite{kir16}.  However, the surviving planetary bodies in the 
SDSS\,1557 system likely had primordial orbits beyond a few AU to escape engulfment. 

At the current steady-state accretion rate, over $10^{17}$\,g of planetary debris has been deposited onto the white dwarf since 
its discovery in 2010.  This is the bare minimum mass of the reservoir, yielding a parent body with diameter greater than 4\,km for 
typical asteroid densities.  This implies that the formation of large planetesimals in the SDSS\,1557 system was not inhibited by the 
relatively massive substellar companion, and that destructive collisions among growing planetesimals were avoided.  

If this polluted and dusty white dwarf is similar to the well-studied larger population, then the parent body of the debris was likely
formed within the snow line of the progenitor system, and thus implying that rocky planet formation was robust in this circumbinary
environment.  These observations therefore support a picture where additional mechanisms can promote planetesimal growth in 
the terrestrial zones of close binary stars, which are predicted to be substantially wider than in planet forming disks around single
stars \cite{var16}.  This interpretation can be confirmed with ultraviolet spectroscopy of key carbon and oxygen transitions, as 
these will easily distinguish between rocky and icy parent bodies.

\bibliographystyle{Sci.}

\begin{figure*}
\begin{center}
\includegraphics[width=150mm]{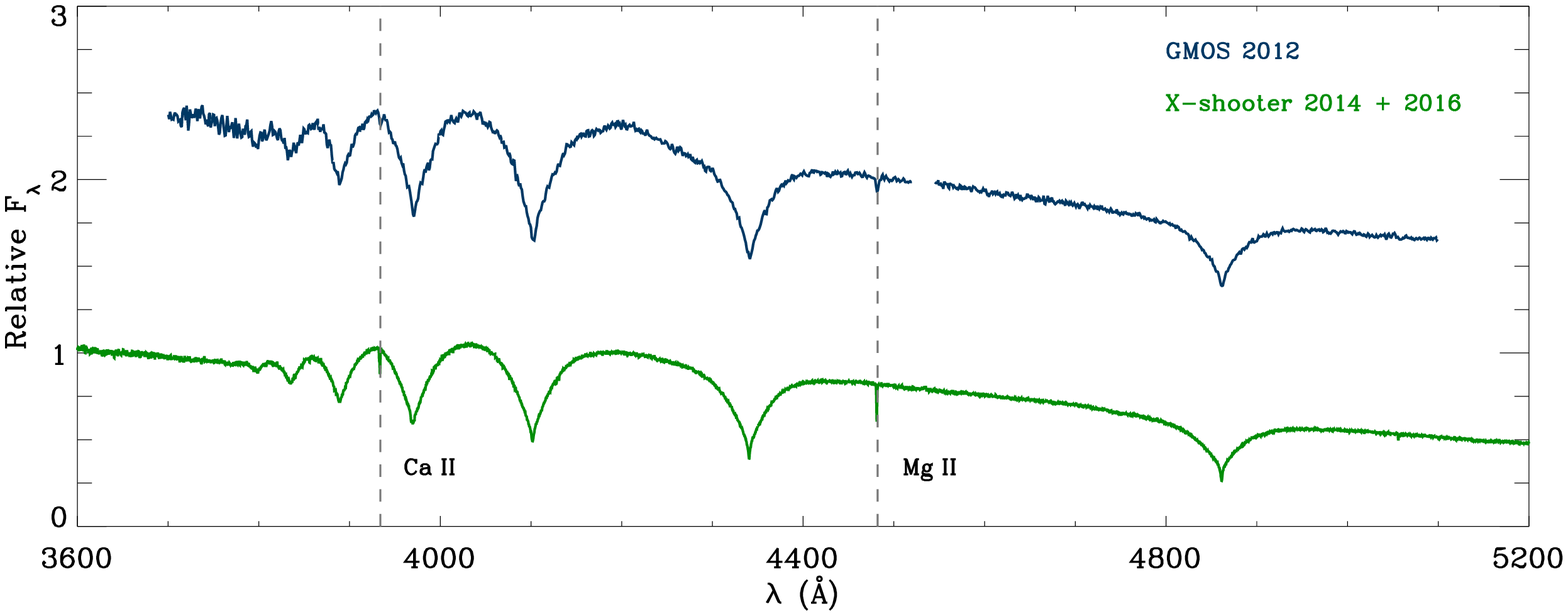}
\includegraphics[width=150mm]{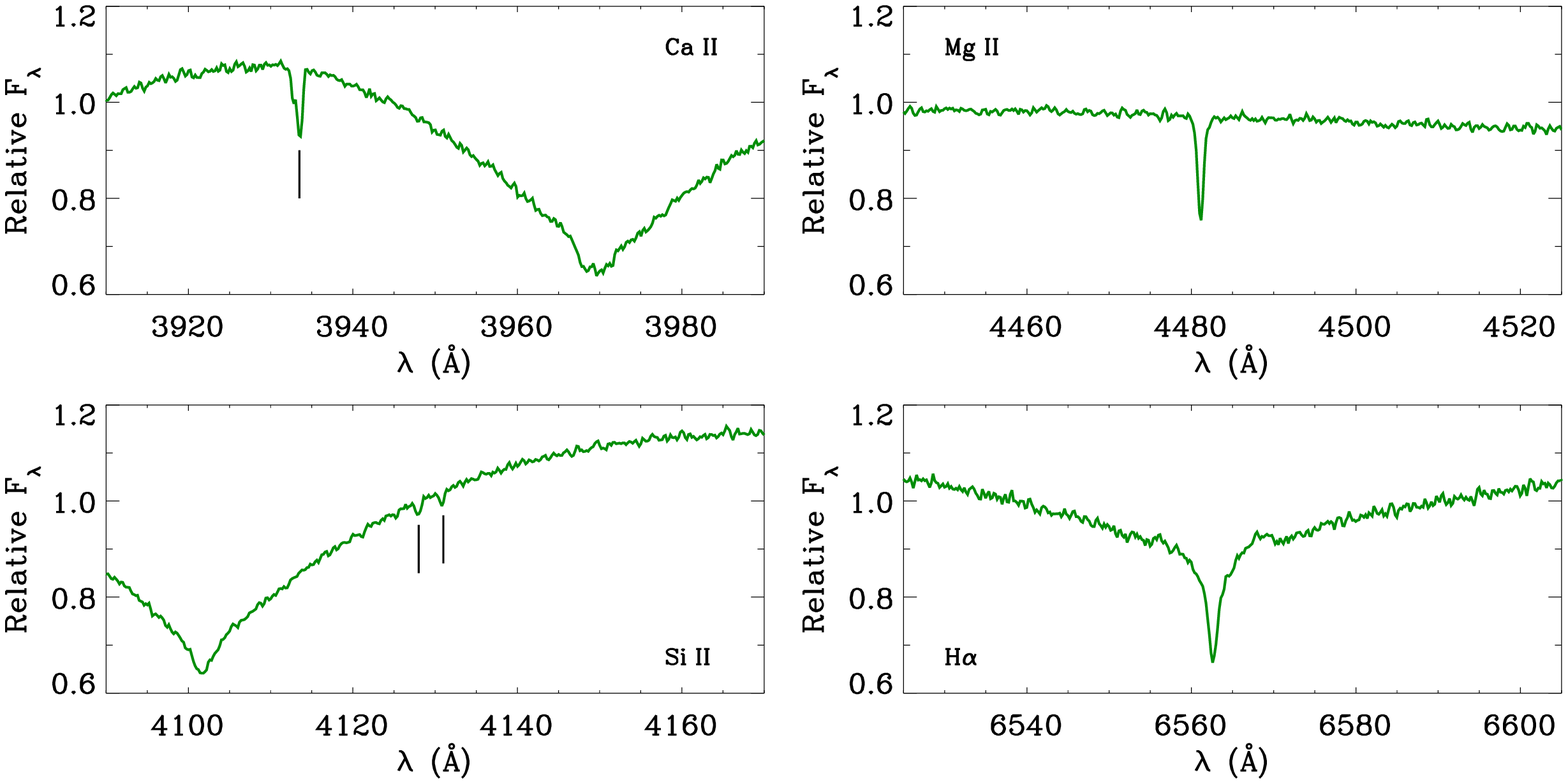}
\end{center}
\vskip -10 pt
{\footnotesize {\bf Supplementary Figure 1: Spectroscopic data plots.}  The top panel shows the combined GMOS data set in blue.  In 
addition to the Mg\,{\sc ii} 4482\AA \ feature detected previously in the ISIS spectrum, the Ca\,{\sc ii} K line is also clearly detected; this 
is unprecedented at low-resolution for a $T_{\rm eff}>20\,000$K, hydrogen atmosphere white dwarf.  In green is the velocity-shifted 
and combined, X-Shooter spectrum over the same wavelength range.  The lower four figures focus on individual line regions 
within the X-shooter data, where notable features include emission components in both H$\alpha$ and Ca\,{\sc ii} K. Because 
these data are co-added in the rest frame of the white dwarf, the weak H$\alpha$ emission from the irradiated companion is 
smeared over a wide velocity range, resulting in the small amount of structure flanking the sharp photospheric absorption.  
Weak lines of Si\,{\sc ii} are visible as shown at 4128, 4131\,\AA, and also (not shown) at 5056, 6347\,\AA.}\end{figure*}

\begin{table}
\begin{center}	

\small
\begin{tabular}{@{}lccc@{}}

\hline

Epoch					&2010.4			&2012.4				&2014.4, 2016.3--2016.6\\
Facility					&WHT			&Gemini South			&VLT\\		
Spectrograph				&ISIS			&GMOS				&X-shooter\\
$\lambda / \Delta \lambda$	&2300			&1600				&6000\\
$\lambda$ (\AA)			&3650--5120		&3610--6450			&3000--10\,2000\\
\# Spectra					&3				&6					&60\\
Detected Metals			&Mg\,{\sc ii}		&Mg\,{\sc ii}, Ca\,{\sc ii}	&Mg\,{\sc ii}, Si\,{\sc ii}, Ca\,{\sc ii}\\
$T_{\rm eff}$ (K)			&$22\,250\pm1190$	&$21\,800\pm600$		&$21\,810\pm790$\\
$\log\,g$ (cm\,s$^{-2}$)		&$7.61\pm0.36$	&$7.50\pm0.10$		&$7.63\pm0.11$\\
$M$ ($M_{\odot}$)			&$0.43\pm0.14$	&$0.40\pm0.04$		&$0.45\pm0.04$\\

\hline

\end{tabular}
\end{center}
\normalsize

{\footnotesize {\bf Supplementary Table 1: Summary of spectroscopic observations.}  In addition to the spectrum available from the 
SDSS,there were three follow up, spectroscopic data sets obtained for SDSS\,1557, whose details are listed in the table.  Within each
data set, the higher Balmer lines for individual spectra were fitted with atmospheric models to obtain $T_{\rm eff}$ and $\log\,g$, 
where the listed values and errors in the table are the resulting means and standard deviations, respectively.  Stellar masses are 
derived from evolutionary models, with errors propagated from the uncertainties in $T_{\rm eff}$ and $\log\,g$.  The approximate 
resolving power $R = \lambda / \Delta \lambda$ is given at 4000\,\AA.  The X-shooter values were adopted for this work, as 
these data are of the highest quality and spectral resolution.}

\end{table}

\begin{table}
\begin{center}

\small
\begin{tabular}{@{}crrccrr@{}}

\hline

HJD			&Radial Velocity	&Error		&&HJD			&Radial Velocity	&Error\\
			&(km\,s$^{-1}$)		&(km\,s$^{-1}$)	&&				&(km\,s$^{-1}$)		&(km\,s$^{-1}$)\\
\hline

56\,798.602714		&43.85		&5.62		&&57\,488.821731		&$-$16.57		&3.27\\
56\,798.620193		&$-$8.53		&4.58		&&57\,488.830599		&$-$10.38		&3.54\\
56\,798.637932		&$-$25.42		&4.92		&&57\,488.839467		&9.88		&4.54\\
56\,798.655365		&3.76		&4.54		&&57\,512.721651		&36.55		&3.31\\
56\,798.672982		&52.03		&4.29		&&57\,512.730518		&58.61		&3.37\\
56\,798.690741		&45.63		&4.77		&&57\,512.739384		&63.01		&3.21\\
56\,798.708633		&6.45		&4.90		&&57\,512.748670		&52.09		&3.19\\
56\,798.739539		&$-$18.84		&4.71		&&57\,512.757559		&29.85		&3.25\\
56\,798.757188		&24.90		&4.73		&&57\,512.766426		&9.63		&3.17\\
56\,798.769488		&41.84		&11.32		&&57\,512.775554		&$-$5.40		&3.02\\
56\,799.679388		&$-$17.17		&5.31		&&57\,512.784445		&$-$9.75		&3.21\\
56\,799.695131		&$-$0.67		&4.96		&&57\,512.793335		&$-$7.41		&3.20\\
56\,799.713580		&40.20		&5.87		&&57\,566.532498		&56.37		&3.55\\
56\,799.729323		&54.91		&8.13		&&57\,566.541386		&55.72		&3.56\\
56\,799.747266		&14.84		&14.33		&&57\,566.550262		&36.20		&3.34\\
56\,799.763003		&$-$29.29		&14.22		&&57\,569.571379		&61.02		&2.89\\
56\,799.781362		&$-$18.10		&11.44		&&57\,569.580259		&55.07		&2.99\\
56\,799.797089		&23.01		&20.21		&&57\,569.589125		&21.26		&3.02\\
57\,486.793931		&58.20		&3.12		&&57\,569.609550		&$-$20.62		&3.06\\
57\,486.802812		&39.27		&3.39		&&57\,569.618431		&$-$22.69		&2.97\\
57\,486.811696		&15.03		&3.46		&&57\,569.627298		&3.82		&3.02\\
57\,486.823164		&$-$15.73		&3.42		&&57\,597.534207		&2.43		&3.79\\
57\,486.832047		&$-$26.81		&3.55		&&57\,597.543096		&$-$8.28		&3.51\\
57\,486.840931		&$-$14.69		&3.32		&&57\,597.551991		&$-$21.79		&3.97\\
57\,486.852416		&10.89		&3.22		&&57\,601.534731		&$-$19.29		&3.03\\
57\,486.861294		&32.82		&3.35		&&57\,601.543618		&$-$9.66		&3.23\\
57\,486.870173		&54.40		&3.40		&&57\,601.552518		&11.86		&3.61\\
57\,488.783416		&65.05		&3.48		&&57\,603.486847		&39.71		&3.89\\
57\,488.792296		&42.14		&3.90		&&57\,603.495741		&21.08		&5.72\\
57\,488.801179		&17.24		&4.05		&&57\,603.504640		&7.37		&4.11\\

\hline

\end{tabular}
\end{center}
\normalsize

{\footnotesize {\bf Supplementary Table 2: Mg\,{\sc ii} 4482\AA \ velocity measurements from the X-shooter data.}  The velocities 
were determined by fitting each absorption profile with a combination of a Gaussian and a straight line, with the error derived 
using the damped least-squares method.}
\end{table}

\end{document}